\documentclass[final,1p,times]{elsarticle} 
\usepackage{graphicx} 
\usepackage{amssymb} 
\usepackage{amsthm} 
\usepackage{lineno} 

\journal{Nuclear Physics A} 
\begin{document} 

\begin{frontmatter} 


\title{Is the ridge formed by aligned jet propagation and medium flow?}

\author{Joshua Robert Konzer ''for STAR Collaboration''}

\address{Physics Department, Purdue University,
West Lafayette, IN 47907, USA
email:jrkonzer@purdue.edu}

\begin{abstract} 
Motivated by the recent observation that the ridge decreases with trigger particle angle ($\phi_s$) relative to the event plane, it is theorized that the ridge is formed by interplay between jet propagation and medium flow.  Such interplay may produce asymmetry in the ridge azimuthal correlation at a fixed $\phi_s$.  We present an analysis of this asymmetry from STAR data.  We found an asymmetric ridge with maximum asymmetry at $\phi_s\approx45^{\circ}$ concurrent with a symmetric jet at all $\phi_s$.
\end{abstract} 

\end{frontmatter} 



{\bf Introduction: }Previous studies on dihadron correlation have revealed long range correlation in pseudo-rapidity$(\Delta\eta)$ for near-side hadrons correlated with high $p_T$ trigger particles with small azimuth  $(\Delta\phi)$~\cite{new1}.  It was further found that the long range $\Delta\eta$ correlation (referred to as the ridge) decreases as trigger orientation w.r.t.~event plane, $\phi_s=\phi_{trig}-\psi_{EP}$, progresses from in-plane to out-of-plane~\cite{theory}.  Motivated by our data, Chiu and Hwa~\cite{QM08} suggested that alignment between jet propagation and medium flow direction, likely to be found for in-plane trigger particles, may be responsible for the ridge; radiated gluons (within a small angle of the parton direction) become thermalized with the medium and combine with medium partons to form the ridge when they are aligned in the same direction. This model, called the Correlated Emission Model (CEM), predicts a measurable asymmetry in the $\Delta\phi$ correlation of the near-side ridge for trigger particles on a single side of the event plane.

\begin{figure}[ht]
\vspace{-0.2in}
\centering
\includegraphics[width=\textwidth]{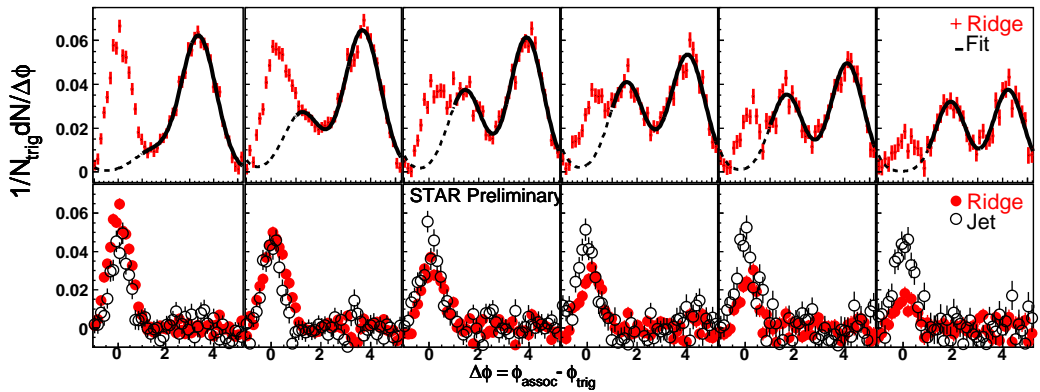}
\vspace{-0.2in}
\caption{
Azimuthal dihadron correlations for both the ridge and jet-like components from in- to out-of-plane in Au+Au 20-60$\%$ collisions.  Trigger and associated $p_T$ ranges are 3$<p_T^{trig}<$4 GeV/c and 1$<p_T^{assoc}<$1.5 GeV/c, respectively.
}
\label{Fig1}
\end{figure}

{\bf Methodology: }The dihadron correlations are separated into $\phi_s$$<$0 and $\phi_s$$>$0 and then further divided into 15$^{\circ}$ slices ranging from in-plane $(\mid\phi_s\mid=0^{\circ})$ to out-of-plane $(\mid\phi_s\mid=90^{\circ})$ for 20-60$\%$ Au+Au collisions.  The trigger particle $p_T$ is 3$<$$p_T^{trig}$$<$4 GeV/c; two associated particle $p_T$ ranges are presented: 1$<$$p_T^{assoc}$$<$1.5 GeV/c and 1.5$<$$p_T^{assoc}$$<$2 GeV/c.  The event plane is constructed from particles outside of the $p_T$ range being analyzed.  For associated particles, the maximum $\langle$$v_2^{2}\rangle$ is calculated by $\langle$cos(2$\Delta\phi$)$\rangle$ of the away-side $\Delta\phi=\phi_{assoc}-\phi_{trig}$ distribution of pairs in the associated $p_T$.  For the trigger, we use $v_2$$\{$RP$\}$ as maximum $v_2$.  For both trigger and associated, we use $v_2$$\{$4$\}$ as a minimum.  The averages of maximum and minimum are used as default $v_2$.  The correlation background shape is given by ~\cite{arXiv}.  The ridge yield is calculated from the background subtracted correlation in $|\Delta\eta|>$0.7.  The systematic uncertainty is bracketed by the max.~and min.~$v_2$ in the background correlation.  To isolate the near-side ridge, the away-side $\Delta\phi$ correlation is fit with a double-Gaussian, same-sigma function,  as shown in Fig.~\ref{Fig1}.  The fit is extrapolated into the near-side and subtracted from the $|\Delta\eta|>$0.7 correlation to remove possible away-side leakage.  The jet-like yield is calculated from the difference between the raw correlations in $|\Delta\eta|<$0.7 and $|\Delta\eta|>$0.7 properly scaled by the $\Delta\eta$ acceptance factor.  Since $v_2$ is independent of $\Delta\eta$ in the measured range, systematic uncertainties cancel.  

\begin{figure}[ht]
\vspace{-0.1in}
\centering
\includegraphics[height=10pc,width=18pc]{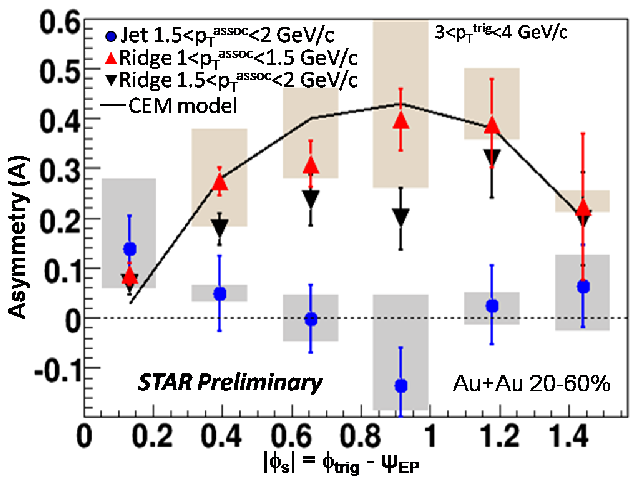}
\vspace{-0.2in}
\caption{
Asymmetry of the ridge and jet-like $\Delta\phi$ correlations as a function of $\phi_s$.  Colored error bars represent statistical uncertainties and shaded bands represent systematic errors due to $v_2$.  Black curve represents CEM prediction.
}
\label{Fig2}
\end{figure}

{\bf{Results: }}As seen in Fig.~\ref{Fig1} and previous studies~\cite{theory}, the ridge decreases with increasing $\phi_s$.  Also noticeable is the apparent asymmetry in the ridge about $\Delta\phi$=0.  Figure ~\ref{Fig2} shows the asymmetry parameter $A = \frac{N_{(0,1)} - N_{(-1,0)}}{N_{(0,1)} + N_{(-1,0)}}$ (where $N_{(a,b)}=\int_{a}^{b}\frac{1}{N_{trig}}\frac{dN}{d\Delta\phi}d\Delta\phi$) as a function of $\phi_s$.  The ridge is found to be asymmetric and the asymmetry persists for different $p_T^{assoc}$ regions.  The maximum asymmetry is found at $\phi_s\approx45^{\circ}$.  The jet asymmetry, shown as comparison, is consistent with zero.  The observed ridge asymmetry is qualitatively consistent with CEM, suggesting that the ridge may be due to jet-flow alignment.  We note that the correlated emission in CEM is rather general, not necessarily restricted to recombination of radiated and medium gluons.  For instance, it is possible that medium fluctuation together with radial flow can produce similar effects~\cite{theories1}.

{\bf Conclusion: }We presented dihadron correlations with respect to the constructed event plane for Au+Au 20-60$\%$ collisions.  The ridge is found to be asymmetric about $\Delta\phi\approx0^{\circ}$ with maximum asymmetry around $\phi_s\approx45^{\circ}$, while the jet remains symmetric about $\Delta\phi\approx0^{\circ}$ within errors.  These results are qualitatively consistent with the Correlated Emssion Model and suggest that the ridge may be formed by aligned jet propagation and transverse radial flow of the bulk medium.



\end{document}